\documentclass{aa}
\usepackage{epsf}

\newcommand{\gcc}{\mbox{~g\,cm$^{-3}$}}
\newcommand{\beq}{\begin{equation}}
\newcommand{\eeq}{\end{equation}}
\newcommand{\req}[1]{Eq.~(\ref{#1})}
\newcommand{\kB}{k_{\rm B}}
\newcommand{\omc}{\omega_{\rm c}}
\newcommand{\omg}{\omega_{\rm g}}
\newcommand{\mel}{m_{\rm e}}
\newcommand{\xr}{x_{\rm r}}
\newcommand{\EF}{\epsilon_{\rm F}}
\newcommand{\Ne}{{\cal N}_B(\epsilon)}
\newcommand{\Necl}{{\cal N}_0(\epsilon)}
\newcommand{\dfde}{{\partial f_0 \over\partial\epsilon}}
\newcommand{\am}{a_{\rm m}}
\begin{document}

  \thesaurus{12.          % A&A Section 12: Physical processes
              (08.14.1;   % Stars: neutron
               02.04.1;   % Dense matter
               02.03.2;   % Conduction
               02.13.1)   % Magnetic fields
             }
% **********************************************************************
\title{Electron conduction in magnetized
       neutron star envelopes}

\author{A.Y.\ Potekhin}
\institute{Ioffe Physical-Technical Institute,
         Politekhnicheskaya 26, 194021 St.~Petersburg, Russia
(palex@astro.ioffe.rssi.ru)
 }
\date{Received  11 May 1999 / Accepted 25 August 1999}
\maketitle
% **********************************************************************
\begin{abstract}
Practical expressions are derived for evaluation of
electrical and thermal conductivities
and thermopower of degenerate electrons
in the outer envelopes of neutron stars with magnetic fields.
All tensor components of the kinetic coefficients are calculated
(those related to conduction 
along and across magnetic field and to the Hall currents).
The kinetic coefficients are presented as
energy averages of expressions containing
energy dependent effective relaxation times of two types,
associated either with longitudinal or with transverse
currents. The calculation
is based on the effective scattering
potential proposed in the previous paper,
which describes the electron-ion and electron-phonon scattering,
taking into account correlation effects
in strongly coupled Coulomb liquid
and multi-phonon scattering in Coulomb crystal, respectively.
Analytic fitting formulae are devised
for the effective relaxation times
at arbitrary field strength.
Basing on these results, we calculate the transport coefficients
at various temperatures, densities, and magnetic
fields pertinent to the neutron star envelopes.

\end{abstract}

\setcounter{page}{787}
% **********************************************************************
%                               TEXT BODY
% **********************************************************************
\section{Introduction}
\label{sect-intro}
Transport properties of neutron star envelopes
determine many aspects of neutron-star evolution.
For instance, the thermal conductivity
in the outer envelope affects cooling of a neutron star and
its radiation spectra
(e.g., Gudmundsson et al.\ \cite{gpe83}; Page \cite{page97};
Potekhin et al.\ \cite{pcy97}).
The electrical conductivity is the basic quantity
for the studies of magnetic-field evolution
(e.g., Muslimov \& Page \cite{mp96};
Urpin \& Konenkov \cite{uk97}; Konar \& Bhattacharya \cite{kb97}),
which in turn affects thermal evolution
(Miralles et al.\ \cite{muk98}).
The thermopower determines a variety of thermomagnetic phenomena
(Urpin \& Yakovlev \cite{uy80'}; Urpin et al.\ \cite{uly86};
Shibazaki et al.\ \cite{shibazaki-ea89}; Yabe et al.\ \cite{ysh91}).

In the outer envelopes
of neutron stars, the transport coefficients are mainly determined
by the processes of electron scattering off strongly correlated ions.
General formalism for calculating kinetic properties of 
strongly coupled Coulomb plasmas
was developed by Hubbard \& Lampe (\cite{hl69})
and Flowers \& Itoh (\cite{fi76})
(see references to earlier results therein).
Yakovlev \& Urpin (\cite{yu80}) derived
approximate analytic expressions, which were
confirmed later in more elaborate calculations
by Raikh \& Yakovlev (\cite{ry82}),
Itoh et al.\ (\cite{itoh-ea83}), and
Nandkumar \& Pethick (\cite{np84}).
Itoh et al.\ (\cite{itoh-ea84,itoh-ea93}) improved
the results of Yakovlev \& Urpin (\cite{yu80})
and Raikh \& Yakovlev (\cite{ry82}) in the solid crust
by taking into account finite sizes of atomic nuclei
(which may be important in the inner crust)
and the Debye--Waller factor
(which describes reduction of electron-phonon scattering rate
due to background
lattice vibrations). The Debye--Waller factor
proved to be important at temperature $T$ close to the
melting temperature of a Coulomb crystal, $T_{\rm m}$,
or at sufficiently high densities
where zero-point lattice vibrations are strong.
Detailed numerical and analytic calculations by
Baiko \& Yakovlev (\cite{bya95,bya96}) were in reasonable agreement
with Itoh et al.\ (\cite{itoh-ea84,itoh-ea93}).

Magnetic fields in the neutron-star envelopes
complicate electron transport making it,
particularly, an\-is\-o\-tropic.
The field strengths of radio pulsars range from $B\sim10^8$~G
to $B > 10^{13}$~G, with typical value $B\sim10^{12}$~G
(Taylor et al.\ \cite{tml93}).
Some X-ray pulsars and soft gamma repeaters are probably magnetars
-- neutron stars with $B\sim10^{14}-10^{15}$~G,
as suggested by Thompson \& Duncan (\cite{td95})
and supported by recent observations
(Vasisht \& Gotthelf \cite{vg97}; Gotthelf et al.\ \cite{gvd99};
Kouveliotou et al.\ \cite{kou-ea98}, \cite{kou-ea99};
Shitov \cite{shitov99}).
Magnetic field strength expressed in the relativistic units,
\beq
   b = \hbar eB/(\mel^2 c^3) \approx B/(4.414\times10^{13}{\rm~G}),
\eeq
is greater than unity for magnetars, unlike for ordinary pulsars.

The magnetic field affects thermodynamic and kinetic properties
of dense degenerate plasmas in different ways,
depending on density, temperature, and field strength
(e.g., Yakovlev \& Kaminker \cite{yk94}).
In general,
electron motion transverse to the field is quantized
into Landau orbitals.
For sufficiently high temperature, however,
the field can be treated as \emph{non-quantizing} (classical).
The non-quantizing magnetic field
does not affect thermodynamic properties
of matter, but hampers transverse transport
and causes the Hall currents.
In a \emph{weakly quantizing} field, where electrons populate several
Landau levels, the thermodynamic functions and kinetic coefficients
oscillate with increasing density around their classical values.
\emph{Strongly quantizing} magnetic field
confines most electrons to the ground Landau level.
In this case, thermodynamic and kinetic properties of matter
are very different from those in the classical regime.

The problem was extensively studied since the end of 1960s
(Canuto \& Ventura \cite{cv77} and references therein).
Expressions for the kinetic coefficients,
based on the solution of the Boltzmann equation
in the relaxation-time approximation
and valid for non-relativistic and relativistic electrons
in the neutron star envelopes with quantizing
magnetic fields were obtained
by Kaminker \& Yakovlev (\cite{kayak}) for transverse transport,
by Yakovlev (\cite{yak84}) for longitudinal transport
and by Hernquist (\cite{hern84}) for all cases.
Potekhin (\cite{p96}, hereafter Paper~I) has shown that
the usual description of electron transport along magnetic field
with the Boltzmann equation yields almost the same
transport coefficients as
a more general kinetic equation for the electron
spin density matrix.
According to Paper~I, the Debye--Waller factor
enhances the longitudinal electrical and thermal
conductivities near $T_{\rm m}$
much stronger if the magnetic field is quantizing.
Basing on these results,
Potekhin \& Yakovlev (\cite{py96}, hereafter Paper~II)
derived practical formulae
for the longitudinal transport coefficients.

The studies cited above
used the customary one-phonon approximation
for the electron scattering in Coulomb solid.
Recently Baiko et al.\ (\cite{baiko-ea98}) have
reconsidered electron transport in non-magnetized plasmas
by including multi-phonon
processes, which have proved to give a contribution of similar magnitude
(but opposite sign) as the Debye--Waller factor near $T_{\rm m}$.
Concerning the Coulomb liquid, Baiko et al.\ (\cite{baiko-ea98}) have
suggested that incipient ordering of ions
in the strong-coupling regime
affects electron scattering; they have proposed
an approximate treatment of this effect by
modification of the static structure factor of ions.
Both modifications (in the solid and liquid phases)
change the kinetic coefficients near the melting point
and drastically reduce their discontinuities at $T=T_{\rm m}$.
The new ion structure factors have been employed by
Potekhin et al.\ (\cite{pbhy99}, hereafter Paper~III)
in calculations of electrical and thermal conductivities
in the outer envelopes of neutron stars without magnetic fields.
Numerical results in Paper~III have been
fitted by analytic expressions,
derived in the relaxation-time approximation
with the use of a specially adjusted
effective scattering potential.

In this paper, the effective potential obtained in Paper~III
is applied
to calculation of electron transport coefficients at arbitrary
magnetic field strength, at temperatures
$T\sim(10^5-10^9){\rm~K}$ and densities
$\rho\sim(10^3-10^{11})\gcc$
typical for the outer envelopes of neutron stars.
Energy-dependent effective relaxation times,
subject to thermal averaging, are obtained
for electron transport parallel and perpendicular
to the quantizing magnetic fields.
In the case of non-quantizing fields, the usual
semiclassical formulae
(e.g., Urpin \& Yakovlev \cite{uy80'}) are utilized
taking into account the results of Paper~III.

The paper is composed as follows.
In Sect.~\ref{sect-basics},
we describe typical
plasma parameters of interest.
In Sect.~\ref{sect-transp},
we express the electron transport coefficients
through an effective scattering potential
in the relaxation time approximation.
In Sect.~\ref{sect-fitPhiPsi},
we present analytic fits to
the effective energy-dependent relaxation times
related to longitudinal and transverse electron transport
in quantizing magnetic fields.
Section~\ref{sect-res} illustrates
the main features of the transport coefficients
given by the present theory.

%=Section  =============================================================
\section{Magnetized degenerate matter}
\label{sect-basics}
% ----------------------------------------------------------------------
Consider a plasma composed 
of electrons (with charge $-e$)
and a single ion species with charge $Ze$
and mass $m_{\rm i}\approx Am_{\rm u}$ 
(where $m_{\rm u}=1.6605 \times 10^{-24}$~g is the atomic mass unit,
and $A$ is the atomic mass number).
The complete pressure ionization occurs at high temperatures or
high densities
(e.g., at $\rho\ga 22\,Z^2 A \gcc$ in the non-magnetic case
-- see Potekhin et al.\ \cite{pcy97}).
Whenever this assumption is violated, we will employ the
mean-ion approximation,
in which all ions in all ionization stages
are replaced by a single species
with some effective values of $Z$ and $A$.
Electrons are assumed to be degenerate and nearly free.
The degeneracy implies $T < T_{\rm F} \equiv (\EF-\mel c^2)/\kB$,
where $\EF$ is the Fermi energy
(including the rest energy, $\mel c^2$)
and $\kB$ is the Boltzmann constant.
Degenerate electrons can be considered as nearly free,
if their kinetic energy
exceeds greatly a typical energy of electron-ion Coulomb attraction;
in a non-magnetized plasma, this happens at $\rho \gg 10 A Z \gcc$
(e.g., Pethick \& Ravenhall \cite{pr95}).

A degenerate electron gas can be characterized by
the Fermi momentum $p_{\rm F}$ or wave number
$k_{\rm F}=p_{\rm F}/\hbar$. Without any magnetic field,
$
k_{\rm F}=k_{\rm F0} \equiv (3\pi^2 n_{\rm e})^{1/3},
$
where $n_{\rm e}\approx \rho Z/(m_{\rm u}A)$
is the electron number density.
It is also convenient to introduce
the ``relativity parameter'' (e.g., Shapiro \& Teukolsky \cite{ST83})
\beq
   \xr = \hbar k_{\rm F0}/\mel c \approx 1.009\,(\rho_6 Z/A)^{1/3},
\eeq
where $\rho_6 \equiv \rho/(10^6\gcc)$.

Let the magnetic field $\vec{B}$ be directed along
the $z$-axis. Then, using the Landau gauge of the vector
potential, the quantum states of a free
electron can be labelled by
the $y$-coordinate of the electron guiding centre,
the longitudinal electron momentum $p_z$,
the Landau number $n$,
and a spin variable $s$.
The ground Landau level ($n=0$) is non-degenerate with respect
to the spin variable ($s=-1$, statistical weight $g_0=1$)
while the other levels ($n>0$) are
doubly degenerate ($s=\pm1$, $g_n=2$).

It is convenient to write (Paper~II)
\beq
    n_{\rm e} = \int_{\mel c^2}^{\infty} \Ne \,
             \left( -\dfde \right)
             {\rm d} \epsilon,
\label{n_e}
\eeq
where $\epsilon$ is the electron energy,
\beq
    f_0(\epsilon) = \left\{ \exp \left[
         (\epsilon - \mu) / \kB T \right]+1 \right\}^{-1}
\eeq
is the Fermi--Dirac distribution,
$\mu$ is the chemical potential ($\epsilon$ and $\mu$
include $\mel c^2$),
and
\beq
    \Ne = {\mel \omc \over 2 (\pi \hbar)^2}
            \sum_{n=0}^{n_{\rm max}} g_n p_n(\epsilon).
\label{N_e}
\eeq
Here and hereafter,
$\omc=eB/\mel c$ is the electron cyclotron frequency,
$p_n(\epsilon)$  is
the value of $|p_z|$ for an electron on the Landau level $n$,
and $n_{\rm max}$ is the maximum Landau number
for a given energy $\epsilon$.
Since
$
 \epsilon = (\mel^2 c^4 + c^2 p_z^2 + 2 \mel c^2 \hbar \omc n)^{1/2},
$
we have
\beq
   p_n(\epsilon) = [(\epsilon/c)^2 - (\mel c)^2
           - 2 \mel \hbar \omc n]^{1/2}
\eeq
and obtain $n_{\rm max}$ as integer part of convenient
(Paper I) dimensionless energy variable
\beq
  \nu = p_0^2(\epsilon) / (2 \mel \hbar \omc).
\label{nu}
\eeq

At $T\ll T_{\rm F}$,
the derivative $(-\partial f_0/\partial\epsilon)$
can be replaced by
the delta function $\delta(\epsilon-\EF)$;
in this case $n_{\rm e}={\cal N}_B(\EF)$.
If $n_{\rm max} \gg 1$, the sum in \req{N_e}
can be approximated by an integral,
which gives the classical (field-free) result,
\beq
   \Necl = p_0^3(\epsilon)/(3\pi^2\hbar^3).
\eeq

Electrostatic screening produced by electrons is
characterized by the Thomas--Fermi wave number $k_{\rm TF}$:
\beq
     k^2_{\rm TF}=4\pi e^2 \, {\partial n_{\rm e}\over\partial \mu}
   = 4\pi e^2
   \int_{\mel c^2}^{\infty} {\partial \Ne \over \partial \epsilon}\,
             \left( -\dfde \right)
             {\rm d} \epsilon.
\eeq

The field can be considered as non-quantizing
if $T\gg T_B$, where (Yakovlev \& Kaminker \cite{yk94})
\beq
    T_B = {\hbar\omg/\kB} \approx
             {1.34\times10^8 (B_{12}/\gamma_{\rm r})}{\rm~K},
\eeq
$B_{12}=B/(10^{12}$~G),
$\gamma_{\rm r}=\sqrt{1+\xr^2}$, and $\omg=eBc/\epsilon$
is the electron gyrofrequency
(for $\epsilon= \EF$).
In the non-quantizing fields,
$T_{\rm F} = T_{\rm r}\,(\gamma_{\rm r}-1)$
and
$
     (k_{\rm TF}/2 k_{\rm F0})^2 \approx
     (\alpha_{\rm f} /\pi)\,\gamma_{\rm r}/\xr,
$
where
$T_{\rm r}\equiv\mel c^2/\kB
\approx 5.93\times10^9$~K and
$\alpha_{\rm f} = e^2/\hbar c \approx 1/137$ is the fine-structure
constant.

The opposite case of strongly quantizing field
occurs at $T\ll T_B$ and $\rho<\rho_B$,
where
\beq
    \rho_B = A m_{\rm u} n_B / Z
          \approx 7.045 \times10^3 \,(A/Z)\,B_{12}^{3/2}\gcc,
\eeq
$n_B=(\pi\sqrt2)^{-1} \, \am^{-3}$,
and $\am = (\hbar c/eB)^{1/2}$
is the so called magnetic quantum length.
In this case,
\beq
    k_{\rm F} = { 2\pi^2 \am^2}\,n_{\rm e}
  = (4/3)^{1/3} ({\rho/\rho_B})^{2/3}
      \,k_{\rm F0}.
\eeq
Therefore $T_{\rm F}$ is strongly reduced
for $\rho \ll \rho_B$.

The state of the one-component plasma (OCP) of
ions depends on the Coulomb parameter,
\beq
  \Gamma = {(Ze)^2\over \kB  T a_{\rm i}}
  \approx {22.75\, Z^2 \over T_6}\left({\rho_6\over A}\right)^{1/3},
\label{Gamma}
\eeq
where
$a_{\rm i}=[3/(4\pi n_{\rm i})]^{1/3}$ is the ion-sphere radius,
$n_{\rm i}=n_{\rm e}/Z$ is the number density of ions,
and $T_6\equiv T/(10^6$~K).
In a weakly coupled OCP, $\Gamma \ll 1$, ions form the Boltzmann gas
whose screening properties are characterized by
the inverse Debye screening length,
\beq
   q_{\rm D} = \sqrt{3\Gamma}/a_{\rm i}.
\eeq
For $\Gamma \ga 1$, the ions constitute
a strongly coupled liquid.
The liquid freezes into a Coulomb crystal
at some $\Gamma = \Gamma_{\rm m}$.
For classical ions (whose zero-point quantum vibrations are negligible),
$\Gamma_{\rm m}\approx175$, whereas
strong zero-point vibrations suppress the freezing and increase
$\Gamma_{\rm m}$ (Nagara et al.\ \cite{nnn87}).
The freezing is completely suppressed in the so called quantum liquids,
which exist at $\xr\ga 0.18 A Z^{7/3}$,
as can be estimated from numerical simulations (Jones \& Ceperley \cite{jc96}).
In general, the quantization of ionic motion is significant
at $T\ll T_{\rm p}$, where
\beq
      T_{\rm p}  =  \hbar \omega_{\rm p} / \kB
          \approx 7.832 \times 10^6 \, (Z/A)\,\sqrt{\rho_6}{\rm~K}
\label{T_p}
\eeq
is the ion plasma temperature,
$\omega_{\rm p}  = ( 4 \pi Z^2 e^2 n_{\rm i} / m_{\rm i} )^{1/2}$
being the ion plasma frequency. We do not consider 
the quantum ion solids and liquids hereafter.

We neglect also effects of magnetic field on
the OCP of ions. This is justified if
the ion cyclotron energy $\hbar\omega_{\rm ci}
=\hbar\omc \, Z\mel/m_{\rm i}$ is small compared with
either $\kB T$ [i.e., $T_6 \gg 0.0737 \,(Z/A)\, B_{12}$]
or typical phonon energies in the OCP
($\sim \hbar \omega_{\rm p}$; i.e., $\sqrt{\rho_6}\,\gg 0.0094 \,B_{12}$).

\begin{figure}
\begin{center}
\leavevmode
\epsfysize=130mm
\epsffile[150 170 460 700]{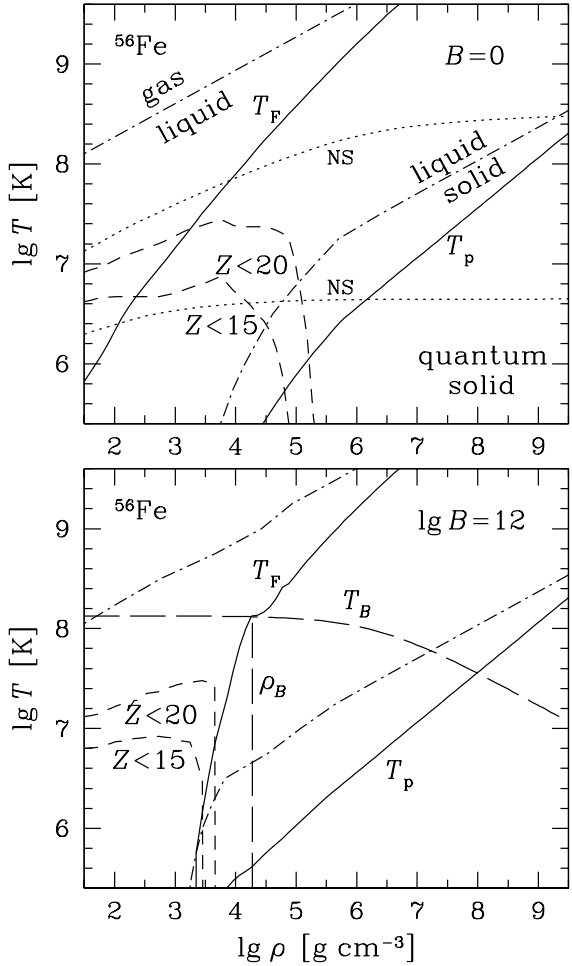}
\end{center}
\caption[]{Characteristic plasma domains on the $\rho-T$ plane for iron.
Upper panel: non-magnetic plasma; lower panel: $B=10^{12}$~G.
Solid lines show $T_{\rm F}$ and $T_{\rm p}$ vs.\ $\rho$;
upper and lower dot-dashed lines correspond to $\Gamma=1$ and $\Gamma=175$,
respectively;
short-dashed lines indicate the domains of partial ionization.
Dotted lines on the upper panel show temperature profiles
in the envelope of a ``canonical'' cooling
neutron star (see text) for two values of
the effective surface temperature,
$2\times10^5$ and $2\times10^6$~K.
Long-dashed lines on the lower panel show $T_B$ and $\rho_B$
and separate the regions
of strong and weak magnetic quantization.
}
\label{fig-domains}
\end{figure}

The characteristic $\rho-T$ domains are shown in Fig.~\ref{fig-domains}
for iron plasma at $B=0$ and $10^{12}$~G.
We have taken into account partial ionization in the mean-ion approximation.
Electrons are degenerate below $T_{\rm F}$;
ions are classical above $T_{\rm p}$;
thus, strictly speaking, our consideration is valid
in the stripe between the solid lines.
In practice it may be reasonably accurate
outside this stripe
because it incorporates thermal averaging
and because the quantum effects are
actually small as long as $T\ga 0.1\,T_{\rm p}$.

The short-dashed contours indicate the region of partial ionization:
the upper contour corresponds to the effective charge $Z=20$
and the lower one to $Z=15$.
We have evaluated $Z$ in the same
manner as Potekhin et al.\ (\cite{pcy97}),
requiring the equation of state (EOS)
of a plasma composed of free electrons and ions
with the effective charge $Z$
to reproduce a ``standard'' EOS
that takes into account partial ionization.
In the non-magnetic case, the OPAL EOS (Rogers et al.\ \cite{OPAL_EOS})
has been adopted as such a standard and,
whenever necessary,
interpolated as explained in Potekhin et al.\ (\cite{pcy97}).
In the magnetic case,
we have used the finite-temperature
Thomas--Fermi EOS by Thorolfsson et al.\ (\cite{Thorolf}).

The dotted curves on the upper panel (marked ``NS'')
reproduce the temperature profiles (Potekhin et al.\ \cite{pcy97})%
\footnote{We have recalculated the profiles using the
updated non-magnetic conductivities (Paper~III), but an effect of the
update turned out to be negligible.}
in the envelope of a ``canonical'' neutron star of the mass
$1.4\,M_\odot$ and radius 10 km,
with an effective surface temperature $2\times10^5$~K
(the lower curve) and $2\times10^6$~K (the upper curve).
In reality, effective temperatures of the middle-aged isolated
neutron stars are believed to lie between these two extremes
(e.g., Page \cite{page98}).

The dot-dashed lines on both panels correspond to
$\Gamma=1$ (gas/liquid smooth transition, upper lines)
and $\Gamma=175$ (liquid/solid phase transition, lower lines).

Finally, the long-dashed lines on
the lower panel indicate three $\rho-T$ regions,
where the magnetic field is strongly quantizing
(to the left of $\rho_B$ and considerably below $T_B$),
classical (much above $T_B$),
or weakly quantizing.

% ----------------------------------------------------------------------
%
\section{Transport coefficients}
\label{sect-transp}
% ----------------------------------------------------------------------
\subsection{General relations}
Consider electron electric and thermal currents
induced in a magnetized plasma
under the effect of an electric field $\vec{E}$ and
weak gradients of chemical potential $\nabla\mu$
and temperature $\nabla T$. 
These currents can be decomposed into conduction
and magnetization components (e.g., Hernquist \cite{hern84}).
The latter ones relate to surface effects and
must be subtracted.
Let $\vec{j}_e$ and $\vec{j}_T$ be the conduction components of
electric and thermal current densities. They can be written as
(e.g., Landau \& Lifshitz \cite{LaLi-ED2})
\beq
   \vec{j}_e = \sigma\cdot\vec{E}^\ast - \alpha\cdot\nabla T,
\quad
\vec{j}_T = \tilde\alpha\cdot\vec{E}^\ast-\tilde\kappa\cdot\nabla T,
\label{J}
\eeq
where $\vec{E}^\ast = \vec{E}+\nabla\mu/e$
is the electrochemical field.
The symbols $\sigma$, $\alpha$, $\tilde\alpha$,
and $\tilde\kappa$ denote second-rank tensors
($\sigma$ is the electrical conductivity tensor)
which reduce to scalars at $B=0$ only.
From the Onsager symmetry relation one obtains:
$\tilde\alpha_{ij}(\vec{B})=T\alpha_{ji}(-\vec{B})=T\alpha_{ij}(\vec{B})$.

Equations (\ref{J}) can be rewritten as
\beq
   \vec{E}^\ast = R\cdot\vec{j}_e - Q\cdot\nabla T,
\quad
   \vec{j}_T = - T Q\cdot\vec{j}_e - \kappa\cdot\nabla T,
\eeq
where $R=\sigma^{-1}$, $Q=-R\cdot\alpha$, and
$\kappa = \tilde\kappa + T\alpha\cdot Q$
are the specific resistance, thermopower, and thermal conductivity
tensors, respectively (here $Q$ is defined
as in Paper~II; an opposite sign has been adopted by 
Hernquist \cite{hern84}).

The components of $\sigma$, $\alpha$, and $\tilde\kappa$ can be expressed as
\beq
\!\left[
   \begin{array}{c}
       \sigma_{ij} \\ \alpha_{ij} \\ \tilde\kappa_{ij}
   \end{array}
   \right]\!
  =
     \int \!\left[
   \begin{array}{c}
       e^2 \\ e {(\mu - \epsilon)/ T} \\ {(\mu - \epsilon)^2/ T}
   \end{array}
   \right]\!
   {\Ne\over\epsilon/c^2} \,\tau_{ij}(\epsilon)
   \!\left( -\dfde \right)
      {\rm d}\epsilon.
\label{sigma-tau}
\eeq
The functions $\tau_{ij}(\epsilon)$
play role of relaxation times
determined by electron scattering
in the magnetic field.
Owing to the symmetry properties of
the tensors $\sigma$, $\alpha$, and $\tilde\kappa$,
there are only three different
non-zero components:
$\tau_{zz}$ related to longitudinal currents,
$\tau_{xx}=\tau_{yy}$ related to transverse currents,
and $\tau_{xy}=-\tau_{yx}$ related to the Hall currents.

% ----------------------------------------------------------------------
\subsection{Non-quantizing magnetic field}
If the quantizing nature of the magnetic field is neglected, then
(e.g., Urpin \& Yakovlev \cite{uy80'})
\beq
   \tau_{zz}=\tau_0,
\quad
   \tau_{xx}={\tau_0\over 1+(\omg\tau_0)^2},
\quad
   \tau_{yx}={\omg\tau_0^2\over 1+(\omg\tau_0)^2},
\label{tau-non-quant}
\eeq
where $\tau_0$ is the non-magnetic relaxation time,
equal to an inverse effective collision frequency
in this case.

In the outer envelopes of neutron stars,
relaxation is mainly determined by electron-ion scattering.
We restrict ourselves to consideration of this mechanism;
possible inclusion of other processes is discussed briefly
in Sect.~\ref{sect-concl}.
In strongly coupled Coulomb plasmas, the scattering is significantly
affected by ion correlations.
In the liquid phase, an appropriate structure factor of ions should
be employed.
In the solid phase, an adequate description is provided
by the formalism of electron scattering off phonons
with allowance for multi-phonon processes
(see Baiko et al.\ \cite{baiko-ea98} for discussion
and references).
In both the liquid and solid regimes, it is convenient
to write the squared Fourier transform of 
the scattering potential as
\beq
     |U_q|^2 =  (4\pi Ze^2)^2 |\phi_q|^2.
\label{Uq}
\eeq
Here, $\phi_q$ is the so called screening function (which
would be equal to $q^{-2}$ for the Coulomb
potential, were the screening neglected).
Then
\beq
   \tau_0(\epsilon) = {p_0^2 v_0^{\phantom{2}} \over
              4\pi n_{\rm i} Z^2 e^4 \Lambda_0(\epsilon)},
\label{tau0}
\eeq
where $v_0 = p_0 c^2/\epsilon$ is an electron velocity, and
$\Lambda_0(\epsilon)$ is the non-magnetic Coulomb logarithm.
In strongly coupled, degenerate Coulomb plasmas,
one has
\beq
    \Lambda_0 = \int_0^{2p_0/\hbar}
    {\rm d}q\,q^3 |\phi_q|^2 S(q) |F(q)|^2\,
    \left[1 -
    \left({\hbar c q\over2 \epsilon}\right)^2  \right],
\label{L}
\eeq
where
$F(q)$ is the form factor of ions,
and $S(q)$ is an effective structure factor that describes
the effects of ion correlations (Baiko et al.\ \cite{baiko-ea98}).

In Paper~III two forms of the effective Coulomb logarithm
($\Lambda_\sigma$ and $\Lambda_\kappa$)
have been obtained from calculations of $\sigma$ and $\kappa$ at $B=0$
beyond the relaxation-time approximation.
The latter approximation
fails if inelastic processes
with energy transfer $\ga \kB T$ are important.
According to Baiko \& Yakovlev (\cite{bya95}) and Paper~III,
this happens at low temperatures $T\ll T_{\rm p}$.
In this paper we will focus on the case
$T\ga T_{\rm p}$, in which
we can neglect the difference
between $\Lambda_\kappa$ and $\Lambda_\sigma$
and adopt $\Lambda_0=\Lambda_\sigma$ for $B=0$.

The transport coefficients due to electron-ion and electron-phonon
scatterings (in the Coulomb liquid and solid, respectively)
can be described with a reasonable accuracy (Paper~III) in a unified way
under the conditions typical for the outer envelopes of neutron stars.
For this purpose it is sufficient to replace
$|\phi_q|^2 S(q) |F(q)|^2$ in \req{L} by
the expression:
\beq
   |\phi_q^{\rm eff}|^2 = {  1- \mathrm{e}^{-w(q)}
        \over (q^2 + q_{\rm s}^2)^2 }
       \, G(t_{\rm p},\beta),
\label{Ufit}
\eeq
where
\begin{eqnarray}
&&\hspace*{-.9em}
     q_{\rm s}^2 = (q_{\rm i}^2 + k_{\rm TF}^2)\,\mathrm{e}^{-\beta},
\label{qs2}
\\&&\hspace*{-.9em}
     q_{\rm i}^2 =
     q_{\rm D}^2\,(1+0.06\,\Gamma) \, \mathrm{e}^{-\sqrt\Gamma},
\label{qi}
\\&&\hspace*{-.9em}
    w(q) = u_{-2} (q / q_{\rm D})^2 \,(1+\beta/3),
\\&&\hspace*{-.9em}
    G(t_{\rm p},\beta) =
  {1+0.122\beta^2\over (1 + t_0^2/t_{\rm p}^2)^{1/2} } D(t_{\rm p}),
\quad
    t_0 = {0.19 \over Z^{1/6}}.
\label{Deta}
\end{eqnarray}
In this case $\mathrm{e}^{-w(q)}$
plays role of an effective Debye--Waller factor
at large $\Gamma$ and is negligible at $\Gamma \la 1$;
$u_{-2}=13$, where $u_j$ is $(\omega/\omega_{\rm p})^j$
averaged over phonon frequencies
$\omega$ in a Coulomb crystal (e.g., Pollock \& Hansen \cite{ph73});
$q_{\rm s}$ is an effective screening wave number;
and $G(t_{\rm p},\beta)$
is a phenomenological factor that describes
reduction of the scattering rate, caused by quantum effects
at $t_{\rm p}\equiv T/T_{\rm p}\ll1$.
The factors $G$, $w$, and $q_{\rm s}$ contain also
phenomenological corrections to the Born approximation
expressed through
$\beta \equiv \pi\alpha_{\rm f} Z p_{\rm F}c/\epsilon_{\rm F}$.
Finally, the function
$
D(t_{\rm p}) = \exp[ - \alpha_0 u_{-1} \,
                         \exp(-9.1 t_{\rm p})/4],
$
where
$
\alpha_0 = 1.683 \sqrt{\xr /  (A Z)}
$
and $u_{-1}=2.8$,
is associated with quantum corrections
to the Debye--Waller factor
(Baiko \& Yakovlev \cite{bya95}).
Note that one can safely set
$G = 1$ for $T\ga T_{\rm p}$ and $Z\la 30$.

% ----------------------------------------------------------------------
\subsection{Transport along quantizing magnetic field}
Let us calculate the longitudinal
electron transport coefficients in the quantizing magnetic field
using the relaxation time
approximation and the effective scattering potential
determined by \req{Ufit}. According to Papers~I and II,
the longitudinal kinetic coefficients
can be written in the form (\ref{sigma-tau})
by defining the effective relaxation time
$\tau_{zz} = \tau_\|$ as
\beq
  {\Ne c^2\over\epsilon} \tau_\|(\epsilon) =
      {(e B)^2 \over 4\pi^3\hbar Z^2 e^4 n_{\rm i}}\,\Phi(\epsilon).
\label{Phi1}
\eeq
The dimensionless function $\Phi(\epsilon)$ is determined by
a kinetic equation for the electron spin density matrix
$\rho_{n s_1 s_2}(z,p_z)$.
It has been shown in Paper~I, however,
that a good accuracy is provided by
a simpler kinetic equation
for the density distribution function
$f_{ns}=\rho_{n s s}$
in the ``fixed spin'' representation (Yakovlev \cite{yak84}).

In order to obtain the longitudinal transport coefficients,
it is sufficient to assume that $\vec{E}^\ast$ and $\nabla T$
are collinear with $\vec{B}$.
The electron distribution function can be sought in the form
\beq
   f_{ns} = f_0 + l\,{\rm sign} (p_z) \dfde
       \left[ e E^\ast + {\epsilon - \mu \over \kB T}\,
       {\partial T \over\partial z} \right] \varphi_{ns}(\epsilon),
\eeq
where $l$ is an appropriate scale length, and $\varphi(\epsilon)$
is a dimensionless function to be determined from the kinetic equation.
The latter is reduced to an algebraic system (Yakovlev \cite{yak84}):
\beq
   \sum_{\gamma n's'} a^{(\gamma)}_{ns\to n's'}(\epsilon)
        \left[ \varphi_{ns}(\epsilon)
           - \gamma\varphi_{n's'}(\epsilon) \right] = 1.
\label{systemPhi}
\eeq
Here, $a^{(\gamma)}_{ns\to n's'}$ is a dimensionless
scattering rate of an electron
from a state with quantum numbers $n$ and $s$
into a state with quantum numbers $n'$ and $s'$, with
changed ($\gamma=-1$) or unchanged ($\gamma=1$)
direction of motion along $\vec{B}$.
The summation is performed over $n'\leq n_{\rm max}(\epsilon)$,
$\gamma = \pm1$, and $s'=\pm1$ for $n'\geq1$
(but $s'=-1$ if $n'=0$).
Note that the terms with $n'=n$, $s'=s$, and $\gamma=+1$
naturally vanish.

Since the scattering potential is written in the form (\ref{Uq}),
it is convenient to choose
$l = \mel c^2 \hbar\omc / (2\pi n_{\rm i} Z^2 e^4)$.
Then (Yakovlev \cite{yak84})
\beq
   \Phi(\epsilon) = \sum_{ns} \varphi_{ns}(\epsilon)
\eeq
and\footnote{%
Equation (\ref{a}) reproduces Eq.\,(26) of Yakovlev (\cite{yak84})
corrected for a misprint.
}
\beq
   a^{(\gamma)}_{ns\to n's'}(\epsilon) =
       [4(\tilde\epsilon+1)^2 \tilde p_n \tilde p_{n'}]^{-1}
      S^{(\gamma)}_{nn'ss'}(\tilde\epsilon),
\label{a}
\eeq
where $\tilde\epsilon=\epsilon/\mel c^2$,
$\tilde p_n = p_n/\mel c$,
\begin{eqnarray}
&&\hspace*{-.9em}
   S^{(\gamma)}_{nn'11} = [(\tilde\epsilon+1)^2
                 + \gamma \tilde p_n \tilde p_{n'} ]^2\, Q_1
\nonumber\\&&\qquad\quad
        + 4b^2nn'Q_2
        + 4b\,[(\tilde\epsilon+1)^2
                 + \gamma \tilde p_n \tilde p_{n'} ]\,\sqrt{nn'}\,Q_3,
\nonumber\\&&\hspace*{-.9em}
   S^{(\gamma)}_{nn',1,-1} =
      2b\,[n'\tilde p_n^2 Q_1 + n\tilde p_{n'}^2\, Q_2 ] - 4b\gamma
         \,\sqrt{nn'}\,\,\tilde p_n \tilde p_{n'}\, Q_3 ,
\nonumber
\end{eqnarray}
and
$S^{(\gamma)}_{nn',-s,-s'}$ differ from $S^{(\gamma)}_{nn'ss'}$
by interchanging $Q_1$ and $Q_2$.
Here the functions $Q_i$ (Yakovlev \cite{yak84})
are generalized to arbitrary scattering potential:
\begin{eqnarray}
   Q_1 & = &
   \int_0^\infty I_{n-1,n'-1}^2(u)\,\tilde\phi^2(u)\,{\rm d}u ,
\label{Q1}
\\
   Q_2 & = &
   \int_0^\infty I_{nn'}^2(u)\,\tilde\phi^2(u)\,{\rm d}u ,
\label{Q2}
\\
   Q_3 & = &
   \int_0^\infty I_{nn'}(u)I_{n-1,n'-1}(u)\,\tilde\phi^2(u)\,{\rm d}u ,
\label{Q3}
\end{eqnarray}
where
\beq
   I_{nn'}(u) = \left({n'!\over n!}\,u^{n-n'}{\rm e}^{-u}\right)^{1/2} 
       L_{n'}^{n-n'}(u)
\eeq
is a Laguerre function
(Sokolov \& Ternov \cite{SokTer};
$L_n^m(u)$ are the associated Laguerre polynomials -- 
e.g., Abramowitz \& Stegun \cite{Abramowitz}),
$\tilde\phi(u) = 2|\phi_q|/\am^2$,
and
$(\hbar q)^2$ is set equal to
$(p_n-\gamma p_{n'})^2+2(\hbar/\am)^2 u$.

\begin{figure}
\begin{center}
 \leavevmode
\epsfysize=83mm
 \epsffile[65 200 550 660]{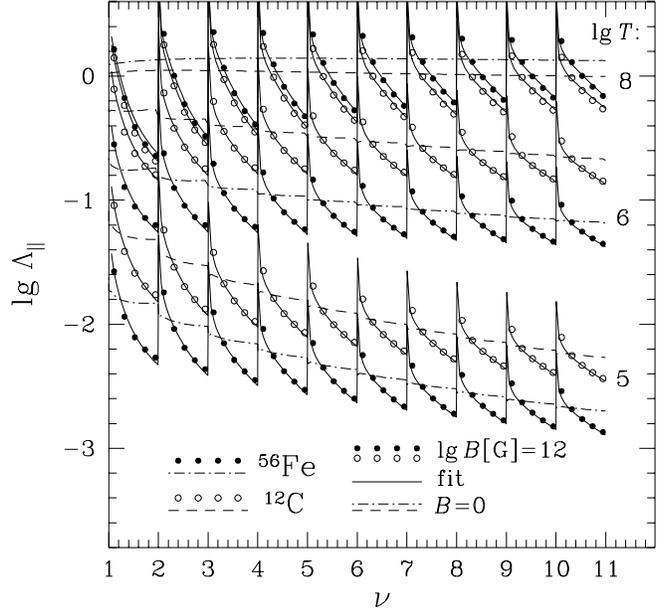}
\end{center}
\caption[]{Calculated (symbols)
and fitted (solid lines) effective longitudinal
Coulomb logarithms $\Lambda_\|$
vs.\ dimensionless electron energy variable \protect{\req{nu}}
for fully ionized iron
(filled circles) and carbon (open circles)
at $B=10^{12}$~G and three values of temperature
($\lg T$ is given at the right end of each bunch of curves).
Dot-dashed and dashed curves represent
the field-free Coulomb logarithms $\Lambda_0$
for Fe and C from Paper~III.
}
\label{fig-b12}
\end{figure}

% ----------------------------------------------------------------------
\subsection{Transport perpendicular to quantizing magnetic field}
Let us start with the case of large Hall parameter,
$\tau_0 \omg \gg 1$.
In this limit, as follows from \req{tau-non-quant},
\beq
   \tau_{yx}\approx1/\omg
\quad{\rm and}\quad
   \tau_{xx}\approx(\omg^2 \tau_0)^{-1},
\label{transv-non-quant}
\eeq
provided the magnetic field is non-quantizing.
The case of quantizing field
was considered
by Kaminker \& Yakovlev (\cite{kayak}) and Hernquist (\cite{hern84}).
The expressions for the transport coefficients
derived by these authors
can be written in the form (\ref{sigma-tau})
by defining, in analogy with \req{transv-non-quant},
\beq
   \tau_{yx}=1/\omg
\quad{\rm and}\quad
   \tau_{xx}=(\omg^2 \tau_\perp)^{-1}, 
\label{tau-large_Hall}
\eeq
where
\begin{eqnarray}
&&\hspace*{-.9em}
  {\Ne c^2\over\epsilon}\,{1\over \omg^2 \tau_\perp(\epsilon)} =
 {Z^2 e^4 n_{\rm i} \over 2\pi\hbar^3 \omc^2}\,\Psi(\epsilon),
\label{Psi1}
\\&&\hspace*{-.9em}
   \Psi(\epsilon) = \sum_{nn'\gamma}
     {b\over 2 \tilde p_n \tilde p_{n'} }
        \left[ (\tilde\epsilon^2+1+\gamma\tilde p_n \tilde p_{n'})
            (Q_1^\perp + Q_2^\perp)
\vphantom{\sqrt{nn'}}\right.\nonumber\\&& \qquad \left.
         + 4b\,\sqrt{nn'}\,Q_3^\perp \right],
\label{Psi}
\end{eqnarray}
and functions $Q_i^\perp$ differ from $Q_i$
[Eqs.~(\ref{Q1})--(\ref{Q3})]
by an additional factor $u$ in each integrand.

In weakly quantizing magnetic field,
$\tau_\perp(\epsilon)$
oscillates around $\tau_0(\epsilon)$;
it can be replaced by $\tau_0(\epsilon)$
in the non-quantizing limit.
This allows us to interpolate
between the regimes of large and moderate-to-low
Hall parameters using the formulae:
\beq
   \tau_{xx}={\tau_\perp\over 1+(\omg\tau_\perp)^2},
\quad
   \tau_{yx}={\omg\tau_\perp^2\over 1+(\omg\tau_\perp)^2}.
\label{tau-quant}
\eeq
Equations (\ref{tau-quant})
relate the effective relaxation times $\tau_{xx}$ and $\tau_{yx}$
to the effective transverse electron
collision frequency $\tau_{\perp}^{-1}$ and
correctly reproduce the known
limits (\ref{tau-non-quant}) (non-quantizing field,
arbitrary Hall parameter)
and (\ref{tau-large_Hall}) (arbitrary field,
large Hall parameter).

In addition, our interpolation (\ref{tau-quant}) of $\tau_{xx}$
eliminates the well known divergency,
that arises from direct substitution of \req{Psi1}
in the integrand of \req{sigma-tau} because
$\Psi(\epsilon)$ turns to infinity at each 
Landau threshold as $(\nu - n_{\rm max})^{-1}$.
Previously $\Psi(\epsilon)$
was truncated at some level, estimated 
by a semi-qualitative analysis of physical
processes that could, in principle, eliminate the divergency,
were they included into the theory
(Kaminker \& Yakovlev \cite{kayak}).
One can show, however, that the derivation
of expressions for the transverse transport coefficients
[equivalent to our Eqs.~(\ref{tau-large_Hall})--(\ref{Psi})]
implied that $\omg\tau_\perp\gg1$.
By correcting relations (\ref{tau-large_Hall})
in case where $\omg\tau_\perp$ is not very large,
\req{tau-quant} ensures finiteness of $\tau_{xx}$,
thus making a truncation unnecessary.

%=Section  =============================================================
\section{Fitting formulae for $\tau_\|(\epsilon)$
and $\tau_\perp(\epsilon)$}
\label{sect-fitPhiPsi}
% ----------------------------------------------------------------------
Explicit expressions of $Q_i$ and $Q_i^\perp$
for the function $|\phi_q^{\rm eff}|^2$ in the form (\ref{Ufit})
are given in the Appendix.
Using them,
we have performed extensive calculations
of $\tau_\|$ and $\tau_\perp$ from Eqs.~(\ref{Phi1}),
(\ref{systemPhi})--(\ref{a})
and (\ref{Psi1}).
The key parameters of the function $|\phi_q^{\rm eff}|$, which enters
these expressions,
--
the Debye--Waller parameter 
$a_{\rm DW} = w(2k_{\rm F0})\approx u_{-2}\,(2k_{F0}/q_{\rm D})^2$
and the Coulomb screening parameter
$a_{\rm s} = (q_{\rm s}/2k_{\rm F0})^2$, --
as well as the magnetic field parameter $b$,
varied independently from $10^{-4}$ to $10^2$.
For any value of $b$, the variable
$\nu$ [related to $\epsilon$ via \req{nu}] varied from 0 to 25,
taking on 5--10 values
over each interval $(n_{\rm max},n_{\rm max}+1)$.

\begin{figure}
\begin{center}
 \leavevmode
\epsfysize=83mm
 \epsffile[65 200 550 660]{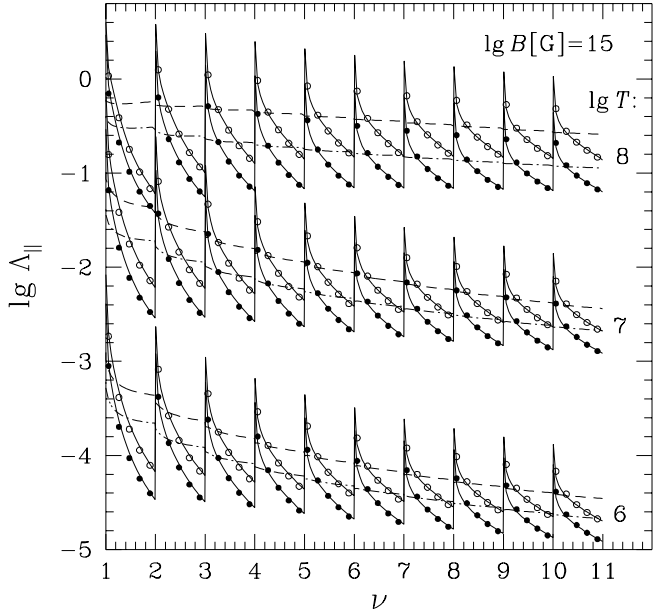}
\end{center}
\caption[]{Same as in Fig.~\protect\ref{fig-b12}
but for $B=10^{15}$~G and different set of $T$.
}
\label{fig-b15}
\end{figure}

\begin{figure}
\begin{center}
 \leavevmode
\epsfysize=83mm
 \epsffile[65 200 550 660]{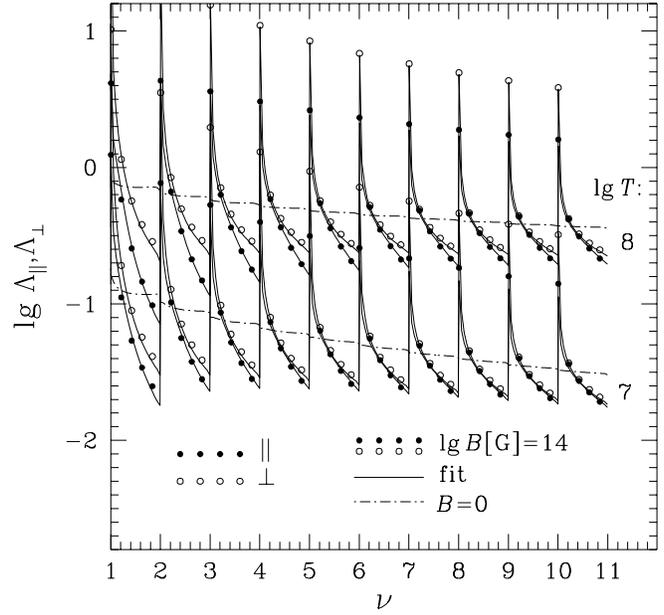}
\end{center}
\caption[]{Longitudinal ($\Lambda_\|$, filled circles)
and transverse ($\Lambda_\perp$, open circles)
effective Coulomb logarithms in iron plasma
for $B=10^{14}$~G and two values of $T$.
The solid lines show the fit; the
dot-dashed lines show $\Lambda_0$.
}
\label{fig-b14a}
\end{figure}

Calculation is quite simple as long as $\nu < 1$,
the functions $\Phi(\epsilon)$ and $\Psi(\epsilon)$
being given by Eqs.~(\ref{Phi2})--(\ref{Q00perp}).
For $\nu > 1$, we have fitted
the results of our numerical calculations by analytic formulae.

Let us define longitudinal and transverse
effective Coulomb logarithms
$\Lambda_{\|,\perp}(\epsilon)$ through the relations
\begin{eqnarray}
   \tau_\|(\epsilon) &=& {\Necl\over\Ne}\,
       {p_0^2 v_0^{\phantom{2}} \over
              4\pi n_{\rm i} Z^2 e^4 \Lambda_\|(\epsilon)},
\label{taupar}
\\
   \tau_\perp(\epsilon) &=& {\Ne\over\Necl}\,
       {p_0^2 v_0^{\phantom{2}} \over
              4\pi n_{\rm i} Z^2 e^4 \Lambda_\perp(\epsilon)}.
\label{tauperp}
\end{eqnarray}
As seen from \req{tau0}, the functions
$\Lambda_\|$ and $\Lambda_\perp$ turn into $\Lambda_0$
if the magnetic field is non-quantizing.
In the quantizing fields, the ratios $\Lambda_{\|}/\Lambda_0$
and $\Lambda_\perp/\Lambda_0$ are fitted by the expressions
\begin{eqnarray}
&&\hspace*{-.9em}\!
   {\Lambda_\|\over\Lambda_0} =
       \left\{ D \left[ 1 + {\sqrt{b}\over\tilde p_0}
           \left( {A\over x} - B\,\sqrt{x}
               + C\,{x-\sqrt{x}\over n_{\rm max} } \right) \right]^{-2}
\right.\nonumber\\&&\hspace*{-.9em}\left.
    +  L^2 x^2 \left[ {3x^2-1\over 2 n_{\rm max} + 1.5 x^2/(1+2b)^2}
              +0.07 + {E\over5} 
          \vphantom{\sqrt{b}\over\tilde p_0} 
\right]^2 
\right\}^{-1/2}
\hspace*{-1.2em},
\label{Lpar}
\\&&\hspace*{-.9em}
   {\Lambda_\perp\over\Lambda_0} =
       1 + {b\over\tilde p_0^2}\,{\tilde A\over x^2}
\nonumber\\&&\hspace*{-.9em}
        +(\sqrt{b}/\tilde p_0)\,
         \left[(\tilde B\,\ln n_{\rm max})\, x^{-1}
            - (\tilde C + \tilde D\,\ln n_{\rm max})\sqrt{x} \right],
\label{Lperp}
\end{eqnarray}
where $x = \tilde p_n/\sqrt{b} = \sqrt{2(\nu-n_{\rm max})}$,
\begin{eqnarray}
&&
    A = {30-15\,E - (15-6\,E)\,v_0^2 \over
            30 - 10\,E - (20-5\,E)\,v_0^2 },
\nonumber\\&&
    B = \frac32 - {E\over2} + \frac14\, {v_0^2 \over 1-2v_0^2/3},
\quad %% \nonumber\\&&
    C = {1-E+0.75\,v_0^2\over 1+v_0^2},
\nonumber\\&&
    D = 1+0.06\,{L^2\over n_{\rm max}^2},
\quad %% \nonumber\\&&
    E = { 1-\exp(-a_{\rm DW})\over a_{\rm DW}} ;
\nonumber\\&&
    \tilde A = 0.8(1+\tilde L)+0.2\,L,
\quad %% \nonumber\\&&
    \tilde B = (0.68-1.3\,\tilde E)\,\tilde L^{1/6},
\nonumber\\&&
    \tilde C = 1.42-\tilde E +\tilde L^{1/2}/3,
\quad %% \nonumber\\&&
    \tilde D = \left(0.52 - \tilde E \right)\tilde L^{1/4},
\nonumber\\&&
    \tilde E = (10+5/b)^{-1} ,
\quad %% \nonumber\\&&
      L = \ln(1+\tilde a^{-1}), \quad
    \tilde L = \tilde a L,
\nonumber\\&&
    \tilde a = \left[
       \sqrt{a_{\rm s}}+(2+0.5\,a_{\rm DW})^{-1} \right]^2.
\nonumber
\end{eqnarray}

\begin{figure}
\begin{center}
 \leavevmode
\epsfysize=85mm
 \epsffile[40 175 550 685]{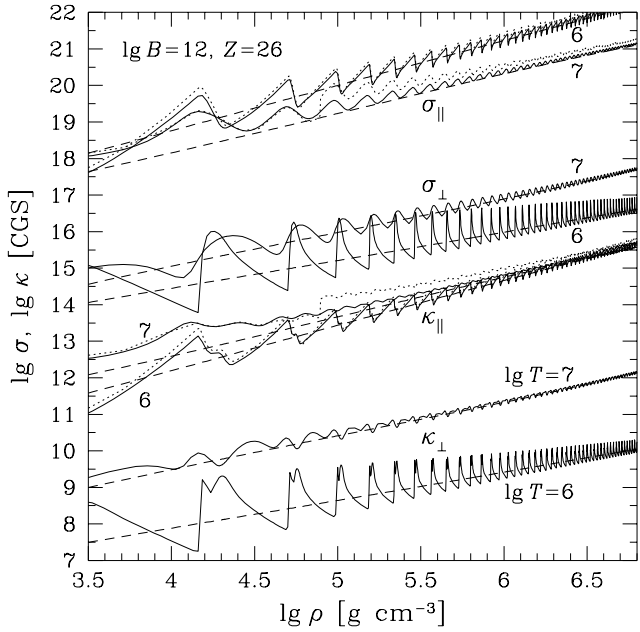}
\end{center}
\caption[]{Longitudinal ($\|$)
and transverse ($\perp$)
electrical ($\sigma$) and thermal ($\kappa$)
conductivities in the outer neutron-star envelope
composed of iron for $B=10^{12}$~G and two values of $\lg T$
(marked near the curves):
comparison of the new results (solid lines)
with the classical approximation
[\protect\req{tau-non-quant}, dashed lines]
and with the old results for the longitudinal
conductivities (dotted lines) from Paper II.
}
\label{fig-z26b12}
\end{figure}

The leading terms at $x\to0$ are proportional to $x^{-1}$ in \req{Lpar}
and $x^{-2}$ in \req{Lperp}, reproducing the asymptotic behaviour 
of the functions $\Phi$ and $\Psi$.
The accuracy of the fit was checked for
$\nu-n_{\rm max}\geq0.01$,
which is quite sufficient for most applications.
Equations (\ref{Lpar}) and (\ref{Lperp}) fit
our numerical results with a typical error of a few
percent. A maximum error up to 40\% occurs only
at some extreme values of $a_{\rm s}$, $a_{\rm DW}$, and $\nu$.
In addition, we have compared the new fit (\ref{Lpar})
with the one in Paper~II
(in the particular case of $a_{\rm DW}\to\infty$
which corresponds to the screened Coulomb scattering potential
considered in Paper~II).
On average, the new fit turned out to be more accurate
than the old one.

Figures \ref{fig-b12}--\ref{fig-b14a} illustrate the accuracy of
Eqs.~(\ref{Lpar}) and ({\ref{Lperp}) for realistic parameters
in the neutron-star envelopes.
In Figs.~\ref{fig-b12} and \ref{fig-b15}, 
the filled and empty circles represent
$\Lambda_\|(\epsilon)$ calculated from Eqs.~(\ref{Phi1})
 and (\ref{taupar});
in Fig.~\ref{fig-b14a} we have additionally plotted
$\Lambda_\perp(\epsilon)$,
obtained using Eqs.~(\ref{Psi1}) and (\ref{tauperp}). The parameters
$a_{\rm s}$ and $a_{\rm DW}$ have been calculated for
fully ionized iron or carbon plasmas
at various $T$ from $10^5$ to $10^8$~K
and $B$ from $10^{12}$ to $10^{15}$~G (indicated in the figures).
At every point, the plasma density has been
determined from the condition $\EF=\epsilon(\nu)$ using \req{nu}.
In all figures,
the fits (\ref{Lpar}) or ({\ref{Lperp}) are drawn by solid lines.
The dashed and dot-dashed lines represent the non-magnetic
Coulomb logarithm $\Lambda_0(\epsilon)$
given in Paper~III.

Since our fitting formulae depend analytically on the parameters
of the effective screening function (\ref{Ufit}),
they need not be changed in case future refinement
of the theory will cause modification of these parameters.

%=Section  =============================================================
\section{Numerical results for transport coefficients}
\label{sect-res}
% ----------------------------------------------------------------------
\begin{figure}
\begin{center}
 \leavevmode
\epsfysize=85mm
 \epsffile[40 175 550 685]{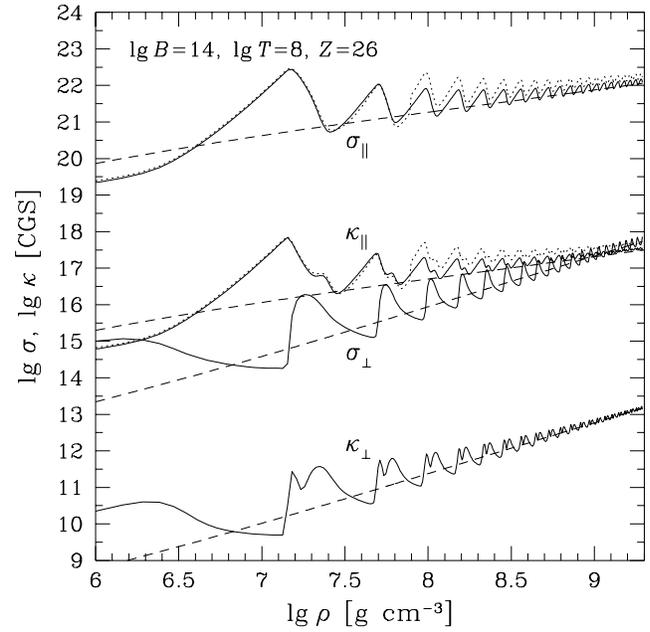}
\end{center}
\caption[]{Same as in Fig.~\protect\ref{fig-z26b12}
but for $B=10^{14}$~G and $T=10^8$~K.
}
\label{fig-z26b14t8}
\end{figure}

Figures \ref{fig-z26b12} and \ref{fig-z26b14t8} show
electrical and thermal conductivities calculated
with the effective relaxation times $\tau_{ij}(\epsilon)$
given by
Eqs.~(\ref{tau-quant})--(\ref{tauperp}) at various $\rho$, $T$, and $B$
appropriate for outer envelopes of the
neutron stars. The use of the
analytic equations (\ref{Phi2})--(\ref{Q00perp}) for $\nu < 1$ and
(\ref{Lpar}), (\ref{Lperp}) for $\nu > 1$
reduces numerical
calculation to one-dimensional integration
in \req{sigma-tau}, which has been performed using a fast algorithm
described in Sect.~5 of Paper~II.

The density range in every figures
allows to see the strongly quantizing (below the first
Landau threshold) and weakly quantizing regimes.
The non-quantizing (classical) results
are plotted by the dashed lines.

Figure \ref{fig-z26b12} shows the longitudinal and transverse
conductivities in a neutron star envelope composed
of iron for $B=10^{12}$~G.
The quantum oscillations around the classical
values are more pronounced at lower temperatures.
Figure \ref{fig-z26b14t8} illustrates the conductivities
at stronger field, $B=10^{14}$~G,
which may be relevant to magnetars.
The classical formulae correctly reproduce the
large-scale trend of the curves and the reduction of
transverse conductivities with respect to longitudinal ones.
Nevertheless, deviations caused by the quantum oscillations
are quite prominent, especially in the regime of strong quantization,
where they may reach orders of magnitude.

\begin{figure}
\begin{center}
 \leavevmode
\epsfysize=85mm
 \epsffile[40 175 550 685]{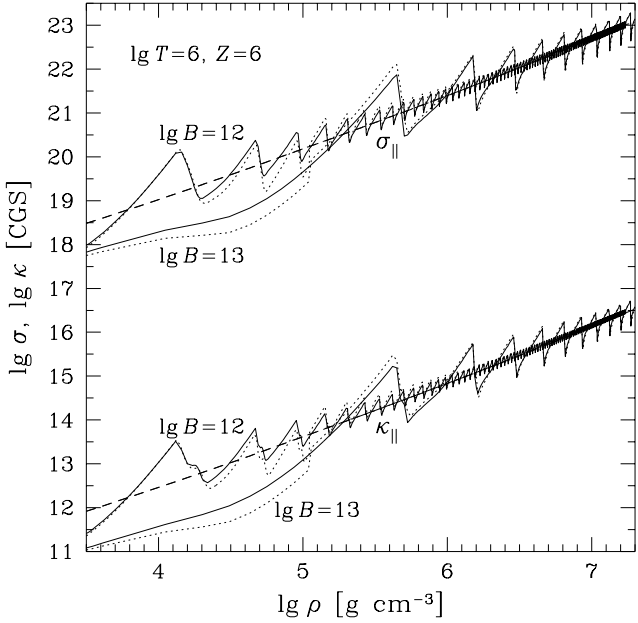}
\end{center}
\caption[]{Longitudinal conductivities
for $Z=6$ and $T=10^6$~K at $B=10^{12}$ and $10^{13}$ G.
As in Figs.~\protect\ref{fig-z26b12}
and \protect\ref{fig-z26b14t8}, the
solid, dotted, and dashed lines show the new, old,
and classical results, respectively.}
\label{fig-z6t6}
\end{figure}

For comparison, we have plotted (by dotted lines)
the longitudinal conductivities
calculated using the formalism of Paper~II.
The temperature values in Fig.~\ref{fig-z26b12}
have been deliberately chosen
the same as in Fig.~5 of Paper~II ($10^6$ and $10^7$~K).
Note that for $T=10^6$~K,
the plasma is entirely within the solid crust.
In this case, our old results agree nicely with the new ones.
On the contrary,
for $T=10^7$~K in Fig.~\ref{fig-z26b12}, as well as for $T=10^8$~K
in Fig.~\ref{fig-z26b14t8},
the displayed density range extends into both the solid crust and
liquid ocean of the star.
The new conductivities go smoothly across the phase transition,
whereas the old ones exhibit large jumps and appear to be
significantly overestimated just behind the ocean/crust interface.
This is caused by an overestimated effect of the Debye--Waller
factor in Paper~II, now corrected by including
multi-phonon processes.

Figure \ref{fig-z6t6} shows the longitudinal conductivities
of carbon plasma
for $B=10^{12}$ and $10^{13}$~G,
which may chance, e.g., in a neutron star with an accreted carbon shell.
In this case, the bottom of the ocean lies slightly above
$\rho=10^5\gcc$. Once again, we observe significant discontinuities
of the ``old" conductivities, which deviate from the new ones
on both sides of the interface. The difference in the liquid
phase is attributed to the modified ion structure factor used
for obtaining $|\phi_q^{\rm eff}|^2$ in Paper~III,
instead of a simplified screened-Coulomb model
in the previous work.

\begin{figure}
\begin{center}
 \leavevmode
 \epsfysize=85mm
 \epsffile[40 175 550 685]{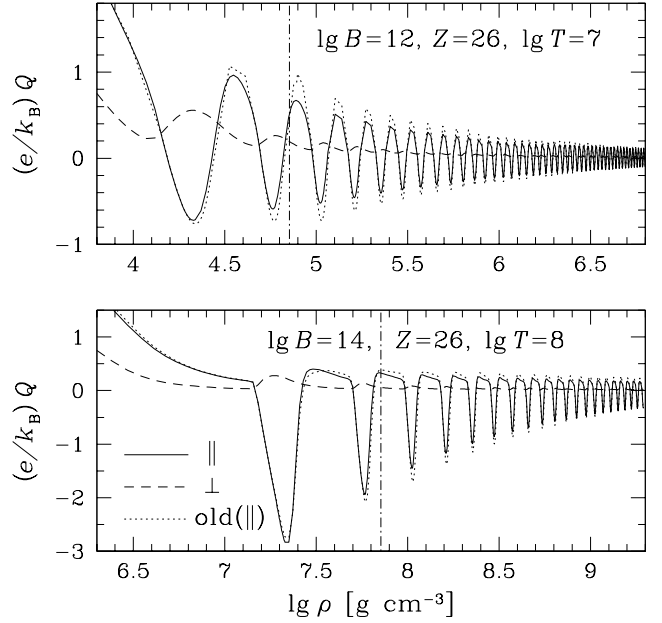}
\end{center}
\caption[]{Longitudinal (solid lines)
and transverse (dashed lines)
thermopower in units of $\kB/e$,
for two sets of indicated parameters.
For comparison, the dotted lines show the
results of Paper~II.
Vertical dot-dashed lines indicate the liquid/solid phase transition.}
\label{fig-Q}
\end{figure}

In Fig.~\ref{fig-Q} we present the
longitudinal and transverse components of the thermopower tensor
for the sets of parameters used in Figs.~\ref{fig-z26b12}, \ref{fig-z26b14t8}.
The longitudinal thermopower from Paper~II is shown by the dotted lines.
Unlike the conductivities, the longitudinal thermopower
did not possess considerable breaks
at the phase transitions in the old theory.
Nevertheless, one can observe that the ``new'' results differ from
the ``old'' ones. As in the previous figures,
this difference is noticeable in the vicinity of the freezing point.

In the previous figures we have presented 
the longitudinal and transverse transport coefficients, which evince
magnetic quantum oscillations around their classical values. 
The off-diagonal (Hall) electrical and thermal conductivities
do not exhibit such oscillations and practically coincide
with their non-magnetic counterparts given by Eqs.~(\ref{sigma-tau}) and 
(\ref{tau-non-quant}). Unlike them, the Hall component
of the thermopower, $Q_{yx}$, does oscillate,
as illustrated in Fig.~\ref{fig-Qhall}
for iron plasma at $B=10^{12}$~G and $T=10^6$, $10^7$,
and $10^8$~K. The oscillations are very sharp at $T=10^6$~K,
but they are completely smeared out 
at the highest temperature, $T=10^8$~K,
which is close to $T_B$ in the present example.
The difference of vertical scales in Figs.~\ref{fig-Q}
and \ref{fig-Qhall}
reflects that $Q_{yx}$ is relatively
small. Nevertheless, it may cause 
a variety of thermomagnetic effects
in neutron star envelopes (Urpin et al.\ \cite{uly86}).

\begin{figure}
\begin{center}
 \leavevmode
 \epsfysize=44mm
 \epsffile[45 270 550 525]{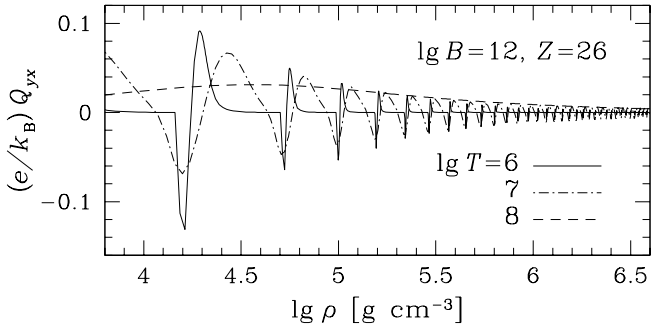}
\end{center}
\caption[]{Hall component of
thermopower in units of $\kB/e$,
for iron plasma at $B=10^{12}$~G
and three values of temperature.}
\label{fig-Qhall}
\end{figure}

%=Section  =============================================================
\section{Conclusions}
\label{sect-concl}
We have derived practical expressions for the electron
transport coefficients in degenerate ($T\la T_{\rm F}$) layers
of neutron-star envelopes with magnetic fields which
may be quantizing. Generally, these expressions require
energy integration, \req{sigma-tau}.
In the case of strongly degenerate electrons
($T\ll T_{\rm F}$) and not too close to the Landau thresholds
($\nu-n_{\rm max}\ga \kB T/ \hbar \omg$),
even this numerical integration becomes unnecessary, and
the electrical conductivity reads
$
   \sigma_{ij} \approx (e^2 n_{\rm e} c^2 /\EF)\,\tau_{ij}(\EF),
$
where $\tau_{ij}$ is provided by our analytic formulae,
while the thermal conductivity is given by the Wiedemann--Franz
law,
$
   \kappa_{ij} \approx (\pi^2\kB^2 T / 3e^2)\,\sigma_{ij}.
$

We have used an effective potential, which 
has been obtained in Paper~III
assuming that the conductivities are determined by
the electron scattering off ions (off phonons in the crystalline phase).
It is well known
(e.g., Yakovlev \& Urpin \cite{yu80}) that
this is the main mechanism regulating electron transport
at $T_{\rm p} \la T \la T_{\rm F}$.
Other contributing mechanisms
are the electron-electron scattering
and scattering off lattice defects and impurities in the crystal.
Corrections due to the impurity scattering can be introduced
in a standard albeit approximate way by summation of relevant
partial collisional frequencies, viz:
$\tau_{\|,\perp}^{-1}=
[\tau_{\|,\perp}^{\rm ei}]^{-1}+[\tau_{\|,\perp}^{\rm imp}]^{-1}$.
In the case of charged impurities
with charge number $Z_{\rm imp}$,
occasionally embedded in a Coulomb lattice,
the effective scattering potential is again given by
Eqs.~(\ref{Uq}) and (\ref{Ufit}),
by setting $G=1$, $\mathrm{e}^{-w(q)}=0$, and $q_{\rm i}=0$
and replacing $Z$ by $|Z-Z_{\rm imp}|$ and $n_{\rm i}$ by $n_{\rm imp}$.
Then
$\tau_{\|,\perp}^{\rm imp}$ are given by our formulae
with an obvious modification of parameters.

We expect that our new formulae for the conductivities
will be useful, in particular, in calculations
of neutron-star thermal structure and evolution.
It would be especially interesting
to apply these results to investigating thermal
structure of magnetars. Up to now,
a very simplified analytic model
(Heyl \& Hernquist \cite{hh-theory}) has been used in this case,
but the problem deserves a more thorough study
since it may provide a clue to the origin of
the anomalous X-ray pulsars (Heyl \& Hernquist \cite{hh97b}).
The formulae presented here are almost as simple as
those used by Heyl \& Hernquist (\cite{hh-theory}), but they are accurate
over a considerably broader range of plasma parameters.

The computer code that implements the formulae
derived in the present paper
is freely available from the author by electronic mail.

\begin{acknowledgements}
I am pleased to acknowledge the hospitality
and financial support of the theoretical astrophysics group
at the Ecole Normale Sup\'erieure de Lyon,
where a part of this work has been done.
I am grateful to D.G.\,Yakovlev for his attention and
useful discussions.
I thank Frank Timmes for a useful advice concerning the computer code.
This work was partially supported by INTAS Grant No.\ 96-542
and RFBR Grant No.\ 99-02-18099.
\end{acknowledgements}

% ====================================================================
\appendix
\addtocounter{section}{1}
\setcounter{equation}{0}
\section*{Appendix: calculation of auxiliary functions}
Let us adopt \req{Ufit} and set $G=1$
for simplicity. Then the dimensionless function $\tilde\phi^2(u)$
that enters Eqs.~(\ref{Q1})--(\ref{Q3})
can be written as
\beq
   \tilde\phi^2(u) = {1-\mathrm{e}^{-\zeta (u+\xi)}
            \over (u+{u_0})^2},
\label{tilde-phi}
\eeq
where $\zeta = 2w(\am^{-1}) = 2 (\am q_{\rm D})^{-2} u_{-2} (1+\beta/3)$,
$\xi = \frac12 (\am/\hbar)^2 (p_n-\gamma p_{n'})^2$,
${u_0}=\xi+\xi_{\rm s}$, and
$\xi_{\rm s} = \frac12 (\am q_{\rm s})^2$.
Let us also define
\beq
   Q^{(0)}_{jnn'm}(\xi,\zeta) =
      \int_0^\infty \! I_{nn'}(u)I_{n-j,n'-j}(u)\,
         {\mathrm{e}^{-\zeta(u+\xi)} \over (u+\xi)^m } \,{\rm d}u.
\label{Q0}
\eeq
Then, for $j=0$ or 1, we have
\begin{eqnarray}
   Q_{2+j} &=& Q^{(0)}_{jnn'2}({u_0},0)
               - \mathrm{e}^{\zeta\xi_{\rm s}}
                 \, Q^{(0)}_{jnn'2}({u_0},\zeta),
\\
   Q_{2+j}^\perp &=& Q^{(0)}_{jnn'1}({u_0},0)
            - {u_0}\, Q^{(0)}_{jnn'2}({u_0},0)
\nonumber\\&&
               - \mathrm{e}^{\zeta\xi_{\rm s}}
               \! \left[  Q^{(0)}_{jnn'1}({u_0},\zeta)
            - {u_0} Q^{(0)}_{jnn'2}({u_0},\zeta) \right],
\label{Qperp}
\end{eqnarray}
and $Q_1, Q_1^\perp$ are obtained from $Q_2, Q_2^\perp$
by replacing $n \to n-1$ and $n' \to n'-1$.

For small Landau numbers ($n,n'\la10$),
one can calculate $Q^{(0)}_{jnn'm}(\xi,\zeta)$
using an explicit expression of the Laguerre functions
$I_{nn'}(u)$ in \req{Q0}.
Since $I_{nn'}(u)=(-1)^{n'-n} I_{n'n}(u)$,
we assume $n'\geq n$ without any loss of generality.
Then
\beq
   I_{n'n}(u) = \mathrm{e}^{-u/2} u^{(n'-n)/2}
       \sum_{k=0}^n (-1)^k c_{n'nk} u^k,
\eeq
where $c_{n'nk} = \sqrt{n'!n!}/[ k!(n-k)!(n'-n+k)!]$.
For $j=0$ or 1,
\begin{eqnarray}
&&\hspace*{-.9em}
   Q^{(0)}_{jn'nm}(\xi,\zeta) = \sum_{l=0}^{2n-j} (-1)^l
          \sum_{k={\rm max}(0,l-n)}^{{\rm min}(n,l)}
        \left[ {n-k \over \sqrt{nn'} } \right]^j
\nonumber\\&&\times
             c_{n'nk} c_{n',n,l-k}
            (n'-n+l)! \, Q^{(0)}_{j,n'-n+l,0,m}(\xi,\zeta),
\label{Q0small}
\\&&\hspace*{-.9em}
   Q^{(0)}_{j,n,0,1}(\xi,\zeta) = (1+\zeta)^{-n}
          \mathrm{e}^\xi E_{n+1}(\xi+\zeta\xi),
\nonumber
\\&&\hspace*{-.9em}
   Q^{(0)}_{j,n,0,2}(\xi,\zeta) = (1+\zeta)^{1-n}
          \mathrm{e}^\xi
             [E_n(\xi+\zeta\xi)-E_{n+1}(\xi+\zeta\xi) ];
\nonumber
\end{eqnarray}
an exponential integral
$E_n(x)=\int_1^\infty t^{-n}\mathrm{e}^{-xt}{\rm d}t$
is easily calculated (Abramowitz \& Stegun \cite{Abramowitz}).

In case where $n$ or $n'$ is large, \req{Q0small} is impractical
because of approximate cancellations of positive and negative terms.
In this case, one can use the following representation%
\footnote{Equation (\ref{B3}) generalizes
Eq.~(B8) of Paper~I to the case in which $\zeta$ and $(m-1)$
may be non-zero simultaneously.}
\begin{eqnarray}
   && \hspace*{-.9em}
   Q^{(0)}_{jn'nm}(\xi,\zeta) = \sum_{k=0}^{n-j}
   {\left[n!\,(n')!\,(n-j)!\,(n'-j)!\right]^{1/2} \over
   k!\,(n-j-k)!\,(n'-j-k)!}
\nonumber\\
   && \times {1\over(m-1)!\,(j+k)!}\,
   \int_\zeta^\infty {x^{2k+j}(x-\zeta)^{m-1}\over (1+x)^{n'+n-j+1}}\,
   \mathrm{e}^{-\xi x}\,{\rm d}x .
\label{B3}
\end{eqnarray}

Finally, let us consider an important particular case
of $n=n'=0$. For transport along and across magnetic field,
the effective inter-collision times $\tau_\|(\epsilon)$
and $\tau_\perp(\epsilon)$ 
are related to the functions $\Phi(\epsilon)$ and $\Psi(\epsilon)$
by Eqs.~(\ref{Phi1}) and (\ref{Psi1}), respectively.
From Eqs.~(\ref{systemPhi})--(\ref{a})
and (\ref{Psi}), we obtain
\beq
   \Phi(\epsilon) = {\tilde p_0^2 \over 2\,Q^\|(\xi) } ,
\label{Phi2}
\quad
   \Psi(\epsilon) = { b\over\tilde p_0^2}
       \left[ \tilde\epsilon^2\, Q^\perp(\xi)
           + Q^\perp(0) \right],
\eeq
where
\begin{eqnarray}
 Q^\|(\xi) & = & {u_0}^{-1}(1-\mathrm{e}^{-\zeta\xi})
        - \mathrm{e}^{u_0} \, E_1({u_0})
\nonumber\\&&
+ (1+\zeta) \, \mathrm{e}^{{u_0}+\zeta\xi_{\rm s}}\,
        E_1({u_0}+\zeta {u_0}),
\\
   Q^\perp(\xi) &=& (1+{u_0}) \,\mathrm{e}^{u_0} \, E_1(u_0) -1
         + \mathrm{e}^{-\zeta\xi}
\nonumber\\&&
-
 (1+{u_0}+\zeta {u_0}) \,\mathrm{e}^{{u_0}+\zeta \xi_{\rm s}} \,
     E_1({u_0}+\zeta {u_0}) ,
\label{Q00perp}
\end{eqnarray}
$\xi=2\tilde p_0^2/b$,
and the function $E_1$ is readily given by polynomial approximations
(Abramowitz \& Stegun \cite{Abramowitz}).

\end{document}